\documentclass[preprint,5p]{elsarticle}
\biboptions{sort}
\usepackage{hyperref}
\usepackage{natbib}
\usepackage{subfig}
\usepackage{color}
\title{The Galaxy Cluster `Pypeline' for X-ray Temperature Maps: ClusterPyXT}

 \author[casa]{Brian Alden\fnref{email}}
 \fntext[email]{brian.alden@colorado.edu}

 \author[casa]{Eric J. Hallman}
 
 \author[casa,nasa]{David Rapetti}

 \author[casa]{Jack O. Burns}
 
 \author[casa,india]{Abhirup Datta}
 
 \address[casa]{Center for Astrophysics \& Space Astronomy, Department of Astrophysical \& Planetary Sciences, 389 UCB, University of Colorado,
Boulder, CO 80309, USA}
 
 \address[nasa]{NASA Ames Research Center, Moffett Field, CA, 94035, USA}
 \address[india]{Indian Institute of Technology, Indore, India}

\newcommand{\msun}{\(M_\odot\)}

\begin{document}
\begin{abstract} \texttt{ClusterPyXT} is a new software pipeline to generate spectral temperature, X-ray surface brightness, pressure, and density maps from X-ray observations of galaxy clusters. These data products help elucidate the physics of processes occurring within clusters of galaxies, including turbulence, shock fronts, nonthermal phenomena, and the overall dynamics of cluster mergers. \texttt{ClusterPyXT} automates the creation of these data products with minimal user interaction, and allows for rapid analyses of archival data with user defined parameters and the ability to straightforwardly incorporate additional observations. In this paper, we describe in detail the use of this code and release it as an open source Python project on GitHub.

\end{abstract}
\maketitle

\section{Introduction}

Galaxy clusters are the largest and most massive ($ \sim 10^{15}$ \msun) gravitationally bound objects in the universe. When clusters of galaxies collide, they produce the most energetic events since the Big Bang, converting as much as $\sim 10^{61}$ ergs from kinetic to thermal energy \cite{galaxy-cluster-formation}. Galaxy clusters contain on the order of hundreds of galaxies in stellar mass, but the bulk of the baryonic mass of the cluster is instead contained in the hot plasma ($10^7 - 10^8$ K) between these galaxies, i.e.~the intracluster medium (ICM) \cite{lowenstein-ICM}. Due primarily to thermal Bremsstrahlung, this high-temperature plasma emits in X-ray, which can therefore be imaged \cite{sarazin} using spaced-based X-ray telescopes such as NASA's {\it Chandra}\footnote{\url{http://chandra.harvard.edu/}} and ESA's {\it XMM-Newton}\footnote{\url{https://www.cosmos.esa.int/web/xmm-newton}}.~These missions have been successfully proving key properties of the ICM in clusters, such as density, mass, and temperature, for a number of years. These observations indicate e.g.~the presence of hydrodynamic shocks resulting from supersonic flows in the ICM caused by major mergers \cite{roettiger-clustershock}. Shocks and turbulence heat and compress the ICM, amplify magnetic fields, and accelerate relativistic particles. The presence of shocks in the ICM is indicated by discontinuities in the X-ray surface brightness and spectral temperature \cite{ryu-shocks}. The software pipeline presented in this paper, \texttt{ClusterPyXT}, automates the creation of high resolution spectral temperature, pressure, and baryonic density maps using X-ray observations of clusters, to facilitate the study of such physical processes.

 The shock-accelerated relativistic particles (electrons) emit synchrotron radiation visible in radio observations \cite{feretti-2012}. These radio relics can be megaparsec\footnote{$1$ Mpc $ \sim 3.086 \times 10^{22}\ $m} (Mpc) sized objects, highlighting the current and past merger history. Located in the periphery of the cluster, these relics can be used to trace shock fronts due to cluster mergers \cite{bruggen-relic-2012} in a complementary manner to that of the X-ray observations.

Moreover, in merging galaxy clusters, radio observations show the existence of radio halos \cite{radiohalo}. These halos, unlike relics, are found in the center of clusters. They may be created by turbulent acceleration or by distributed small-scale shocks \cite{dolag}. The presence of a radio halo or radio relic with steep spectrum is widely accepted as clear evidence of recent merger activity \cite{ryu-shocks}\cite{ensslin-relic-shock}\cite{skillman-2011, skillman-2013}. To probe the physics of merging clusters, it is useful therefore to compare radio observations to the X-ray data products produced by our pipeline.

\subsection{Cluster mergers and shock signatures}
\begin{figure}
	\centering
	\includegraphics[width=\columnwidth]{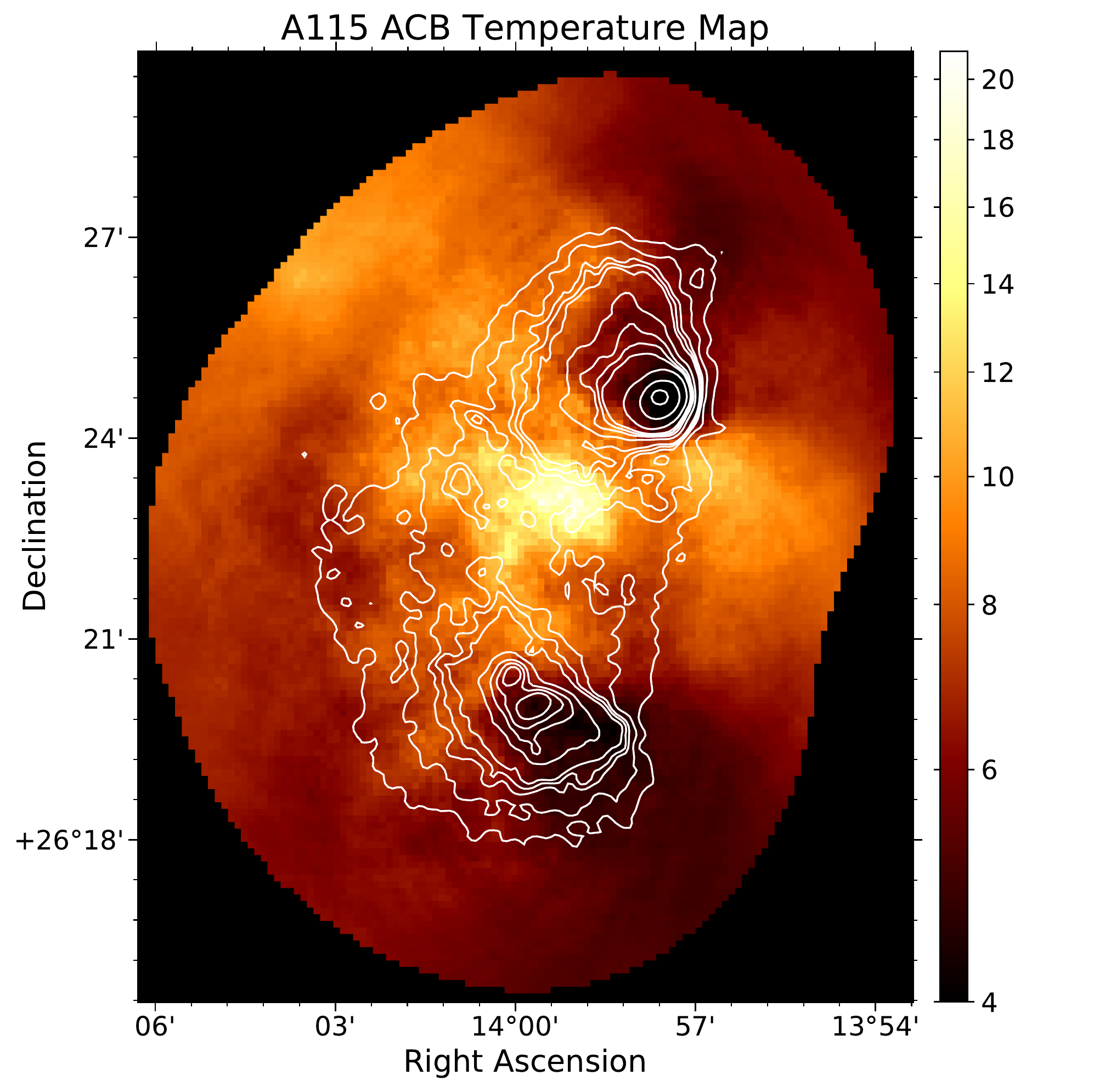}
    \caption{Adaptive circularly binned temperature map of Abell 115. White contours are representative of the X-ray surface brightness. Contour levels are 1, 1.4, 1.8, 2, 2.2, 3.2, 4, 4.6, 5.2, 8, and 20 percent of peak emission ($5.04\times10^{-6}$ counts/cm$^2$/s). The merging subclusters are seen as peaks in the north and south of the X-ray surface brightness. The heated regions in front of each of the cool cores are indicative of the respective directions of travel with respect to the plane of observation. Also of interest is the hot central region, potentially caused by turbulence driven by a bow-shock collision \cite{hallman-a115}. The observation IDs used in this image are 3233, 13458, 13459, 15578, and 15581.}
    \label{fig:a115_tmap}	
\end{figure}

Clusters merge at mildly supersonic speeds. This supersonic infall leads to shock fronts and disruption of the ICM. The resulting shocks heat and compress the gas \cite{burns-stormyclusters}. The disrupted ICM is revealed as non-symmetrical X-ray emission in observations, and discontinuous temperature jumps at the location of the shock features. Temperature maps from these observations can highlight all of this activity. The disturbed ICM will show non-symmetrical hot and cold regions indicative of a ``sloshing'' ICM that is not in hydrostatic equilibrium \cite{tde}. Furthermore, the shock fronts are highlighted by heated regions in front of the subclusters sky-plane projected direction of travel. An example of merger activity heat signatures can be seen in the temperature map shown in Figure \ref{fig:a115_tmap}.

Temperature maps help illuminate the merger process in galaxy clusters.~Looking at the temperature map of A115 in Figure \ref{fig:a115_tmap} it is readily apparent that there are three distinct features.~First, the two peaks of X-ray emission shown in white contours, which appear as cool regions on the temperature map. Second, also immediately apparent is the heated ICM in between the two subclusters. This region could be overlooked in the X-ray image as it is an area of low emission due to the low density. This seemingly uninteresting region in the image, however, becomes a focus of attention in the spectral temperature map. In A115, this heated region tells the tale of turbulence via bow-shock collisions according to the modelling of \cite{hallman-a115}.~Studying cluster X-ray temperature maps in combination with radio observations may also illuminate the physics of shock acceleration, helping us to probe the origin of radio halos and relics. Therefore, high-resolution X-ray temperature maps of galaxy clusters, a central product of the pipeline, are powerful tools for guiding our understanding of the physics taking place in these massive and dynamic objects.

\subsection{High resolution X-ray temperature maps}

There are multiple methods for developing temperature maps using X-ray observations, from which we can calculate the temperature present as the photon energy is recorded in each event. Spectral fitting of these events produces a temperature value for the region fit. \texttt{ClusterPyXT} uses the Astrophysical Plasma Emission Code (\texttt{APEC}) model for optically thin collisionally ionized hot plasma combined with a photoelectric absorption model (\texttt{PHABS}) \cite{APEC-2001,phabs-paper}. These models include redshift, metallicity, temperature, normalization and hydrogen column density. The energies of each photon and the corresponding counts at each energy are fit to the data with this model using the temperature and normalization as the free parameters.~While this aspect of calculating the temperature is standard\footnote{\url{http://cxc.harvard.edu/sherpa/threads/pha_intro/}}, the way the maps are binned is not. Recent findings regarding A2256, the ``Toothbrush'' cluster \cite{van2016lofar-toothbrush}, and A115 \cite{hallman-a115} to name a few, all have featured X-ray temperature maps using a variety of low-resolution binning methods. Our recent work on A115 indicates that high-resolution temperature maps can reveal heretofore unidentified X-ray features that are critical to interpreting the dynamics of merging galaxy clusters \cite{hallman-a115}. \texttt{ClusterPyXT} uses an adaptive circular binning (ACB) \cite{randall-2008, randall-2010, schenck-a85, abhi-a3667} to create a bin for each pixel. The bin size is determined by what size circle, centered on the pixel, is needed to obtain the desired signal to noise ratio. This method creates a bin for each pixel leading to the creation of high-resolution temperature maps. \texttt{ClusterPyXT} automates this process and is freely available as an open source project, coded in Python, on GitHub\footnote{\url{https://github.com/bcalden/ClusterPyXT}}.

While \texttt{ClusterPyXT} was developed for processing X-ray observations of galaxy clusters, the pipeline can also be used to analyze other sources (e.g. galaxies, supernova, active galactic nuclei, etc.). Minor modifications may be required for specific analyses; however, the majority of the underlying code is generalized for any X-ray observation.

\section{ClusterPyXT}

\subsection{Software Overview}

The amount of data generated by the current crop of X-ray telescopes has become significantly large. Many clusters present in the Chandra Data Archive contain more than a hundred kiloseconds of exposure time spread across 10 or more individual pointed X-ray observations.~While existing data processing tools allow for detailed manipulations, automatically chaining these tools together affords a vast reduction of unnecessary manual interaction. The idea of spectral fitting is not new.~Automating this process in an easy to use, open source, and freely available software pipeline is new. There are various types of galaxy cluster X-ray temperature maps that have been described in the literature. While the procedure for spectral fitting is generally the same, the main difference is the binning algorithm used. Various binning algorithms exist, such as weighted Voronoi tesselation (WVT) \cite{wvt}, contour binning \cite{contour-binning}, and adaptive circular binning \cite{randall-2008}\cite{randall-2010}\cite{schenck-a85}. WVT and contour binned maps are generally lower resolution than the corresponding ACB maps. Further, WVT and contour binning assume temperature features align with surface brightness features, but this is not always the case \cite{schenck-a85}. For these reasons, \texttt{ClusterPyXT} uses the ACB algorithm in the spectral fitting routine. Alternate binning methods might be added to the pipeline in a later release.

The temperature maps produced by the pipeline should be used as a qualitative measure of the thermal dynamics of the cluster. Areas of interest in the temperature map should be followed up with targeted analysis for quantitative results. While the temperature map is not the only data product produced (see  \S \ref{sec:data_products}), it is the main deliverable.

\subsection{Program flow}

\begin{figure*}
    \centering
    \includegraphics[width=\textwidth]{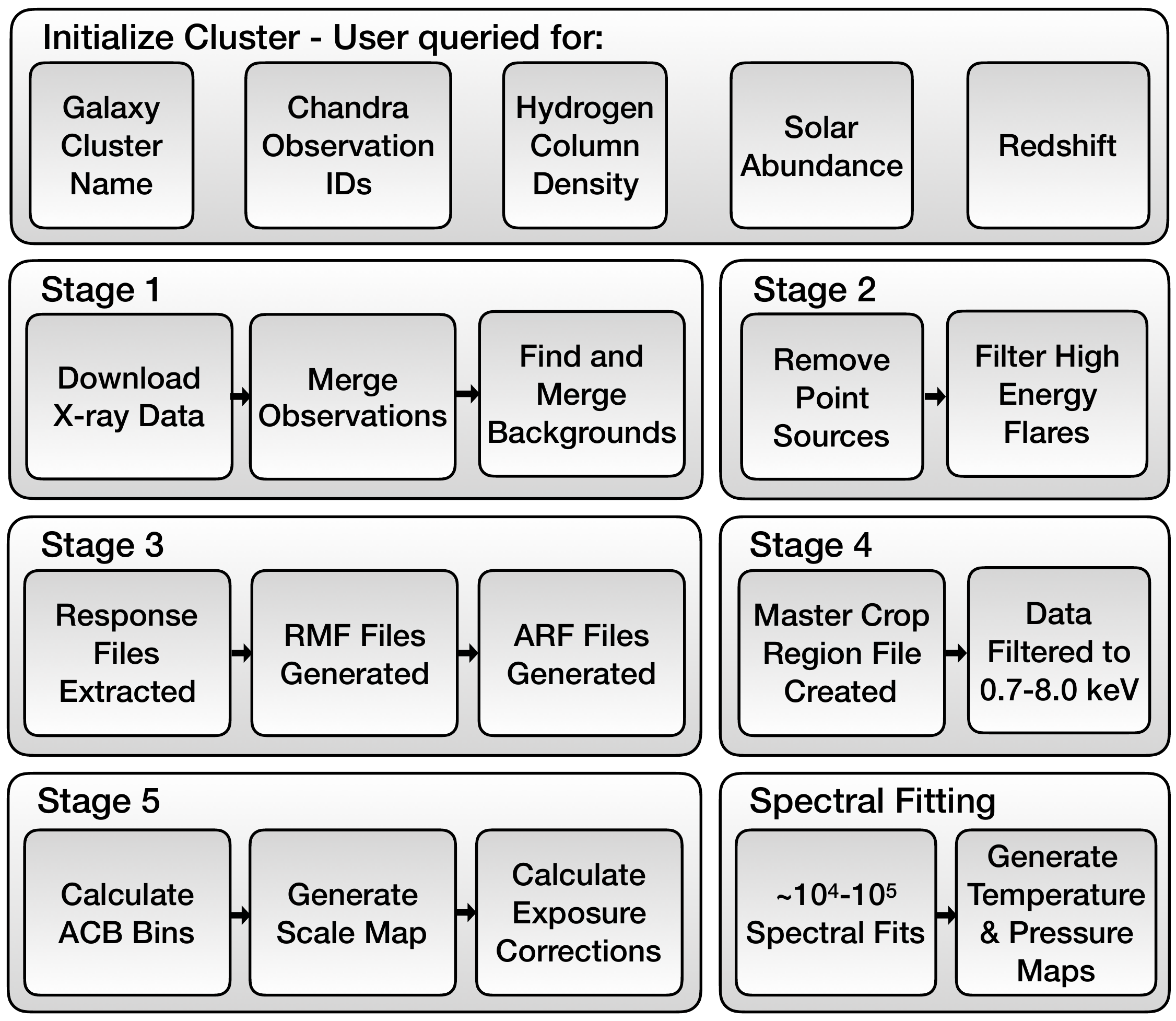}
    \caption{Block diagram showing the program flow.~The overall process consists of cluster initialization (see Section~\ref{sec:key_stages:init}), data acquisition (Stage 1; see Section~\ref{sec:key_stages:1}), data processing (Stages 2-5; see Sections~\ref{sec:key_stages:2}-\ref{sec:key_stages:5}), and data product generation (spectral fitting; see Sections~\ref{sec:spec_fit}~and~\ref{sec:data_products}).}
    \label{fig:flow_diagram}
\end{figure*}

\texttt{ClusterPyXT} automates the process of merging observations and creating the various data products, including temperature maps. First, the user identifies key information about the cluster: name, metallicity, redshift ($z$), Chandra observation IDs (OBSIDs), and the galactic hydrogen column density ($N_H$) along the cluster line of sight. The software proceeds to download the data and merge the observations and corresponding backgrounds. Next, the pipeline removes user provided point source regions followed by filtering high energy events. Following the filtering, the effective exposure times are calculated and a scale map generated, which is effectively the adaptive bin size for each pixel. After this step, the spectral fitting will be completed and a temperature map created. This program flow is visualized in Figure \ref{fig:flow_diagram} and described in detail in Section \ref{sec:key_stages}.

\subsection{Requirements}
The main requirement for \texttt{ClusterPyXT} is the Chandra Interactive Analysis of Observations\footnote{\url{http://cxc.harvard.edu/ciao/}} (\texttt{CIAO}), along with the full Chandra Calibration Database\footnote{\url{http://cxc.harvard.edu/caldb/}} (\texttt{CALDB}) \cite{ciao-caldb}. \texttt{CIAO} needs to be at least version 4.9 with Python 3. While Python 3 is the default distribution installed with \texttt{CIAO}, this is only as of the latest version. For prior versions, the user needs to explicitly indicate Python 3 during the installation process. \texttt{CIAO} is a suite of software tools running in a custom command line environment. It is built and actively maintained by the Chandra X-ray Center\footnote{\url{http://cxc.harvard.edu/}}. This software relies heavily on the operations these tools provide such as response file extraction, reprojections, exposure correction, and the spectral fitting. The calibration files needed for this processing are maintained within \texttt{CALDB}. 

During the installation of the tools in \texttt{CIAO}, it is also required to install the modelling and fitting package \texttt{Sherpa}. This software includes the models used later in fitting, \texttt{PHABS}\footnote{\url{https://heasarc.gsfc.nasa.gov/xanadu/xspec/manual/XSmodelPhabs.html}} \cite{phabs-paper} and \texttt{APEC}\footnote{\url{https://heasarc.gsfc.nasa.gov/xanadu/xspec/manual/node134.html}} \cite{APEC-2001}. The \texttt{PHABS} model is an absorption model for the foreground galactic neutral hydrogen, while the \texttt{APEC} model describes the thermal plasma continuum and line emission. 
\subsection{Usage}
\texttt{ClusterPyXT} can be used through either the command line interface (CLI) or editing configuration files and running the program with command line arguments. While editing configuration files and utilizing command line arguments allows for a more customized pipeline, the remainder of this document will assume the reader is using CLI unless otherwise noted.

\section{Key Stages}\label{sec:key_stages}
\subsection{Cluster Initialization}\label{sec:key_stages:init}
The first step of \texttt{ClusterPyXT} is initializing the galaxy cluster to be analyzed. After starting \texttt{ClusterPyXT} and selecting \texttt{Initialize Cluster} the user is prompted to input the clusters details. These include its name, the observation IDs (currently only Chandra observations with a roadmap to XMM-Newton; see Section~\ref{sec:xmm}), hydrogen column density, redshift, and chemical abundance relative to solar for the cluster. While the user can bypass entering the values for $N_H$, $z$, and abundance at the start, they must be input before completing the spectral fitting. After inputting the required data the next stages of \texttt{ClusterPyXT} can begin.

\subsection{Pypeline Stage 1}
\label{sec:key_stages:1}

The first stage of the pipeline after a cluster is initialized is the initial data acquisition and processing phase. The Chandra observation IDs (OBSIDs) are input to the \texttt{CIAO} \texttt{download\_chandra\_obsid}\footnote{All \texttt{CIAO} commands are documented at \url{http://cxc.harvard.edu/ciao/ahelp/index.html}} function. This function takes these input OBSIDs and downloads all of the files associated with each observation. The current iteration of \texttt{ClusterPyXT} only makes use of the ACIS-I CCDs with ACIS-S processing coming in the next major iteration. After downloading the data for each observation, the data are decompressed.  Each observation is then reprocessed using the \texttt{chandra\_repro} function. This function generates the bad pixel file, level 2 event file, and level 2 pulse height amplitude (PHA) file used in later stages of the pipeline. After reprocessing, for each ACIS-I CCD used in each observation, a level 2 event file is generated using the \texttt{dmcopy} command with the CCD ID specified. These files will then be used with the background files generated in the next step to make a combined X-ray surface brightness map of the cluster.

The background files for each ACIS-I CCD used in each observation are needed to properly obtain a temperature fit. These files are retrieved using the \texttt{acis\_bkgrnd\_lookup} function, and the corresponding backgrounds are reprojected using the \texttt{reproject\_events}\ function, which needs the aspect solution and good time intervals (GTI) for each observation. The aspect solution, the file(s) describing the orientation of the telescope with respect to time, are the files with the \texttt{pcad} prefix. The GTIs, the table of start and stop times for the events, are queried from each of ACIS-I CCD level 2 event files generated in the previous step. The background files are then merged using the \texttt{dmmerge} command.  

\begin{figure*}
    \centering
    \includegraphics[width=\textwidth]{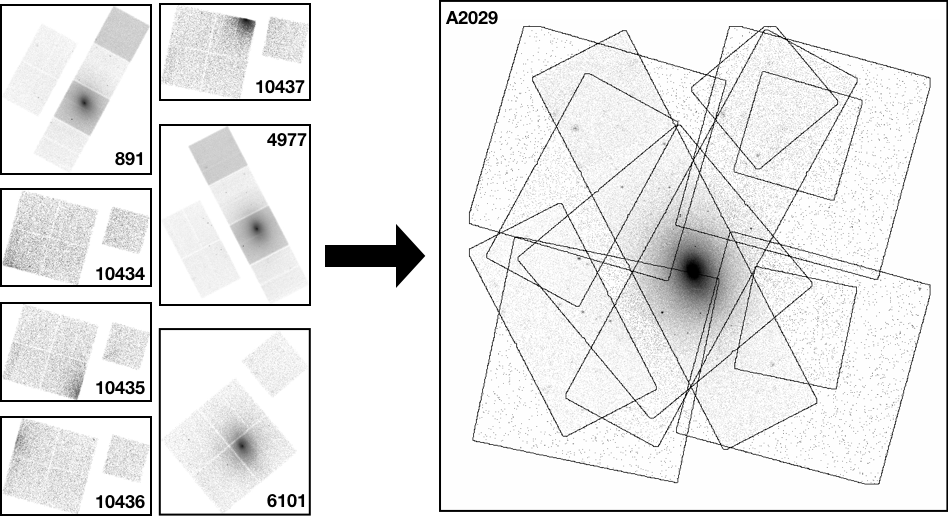}
    \caption{Stage 1 of the pipeline downloads and then merges cluster observations and backgrounds as shown for this example with A2029. Note: The current implementation of the code only utilizes ACIS-I observations. Integration of ACIS-S observations will occur in the next major release.}
    \label{fig:stage_1}
\end{figure*}

After background merger, the pipeline finishes stage one by merging the observations into a combined event file and by creating exposure corrected images. These steps are completed using the \texttt{merge\_obs} and \texttt{fluximage} commands. This stage is illustrated in Figure~\ref{fig:stage_1}. The user is then prompted as to how to proceed using the pipeline.

\subsection{Pypeline Stage 2}
\label{sec:key_stages:2}

After the pipeline merges the event files and creates exposure corrected images into a combined X-ray surface brightness image ([cluster name]\_broad\_flux.img) sources need to be removed. The user is expected to analyze the surface brightness file and create a DS9 region file containing the locations of the sources that are not part of the cluster. The sources region file is passed to the \texttt{dmcopy} command as an exclusion file and a new flux image is created without point sources. 

After source removal, the pipeline filters out time intervals containing high energy background flares from the event files. Flares can be well approximated as a Gaussian. While the bulk of the flare is filtered out during the data analysis, the wings of the distribution often end up in the 2-8 keV range. As such, they still need to be removed beyond just filtering out the energy range. This is completed using the \texttt{deflare} command and passing it high energy lightcurves generated using the \texttt{dmextract} function. This function requires a high energy event file created using \texttt{dmcopy[energy 9000:12000]}.  

\begin{figure*}
    \centering
    \includegraphics[width=\textwidth]{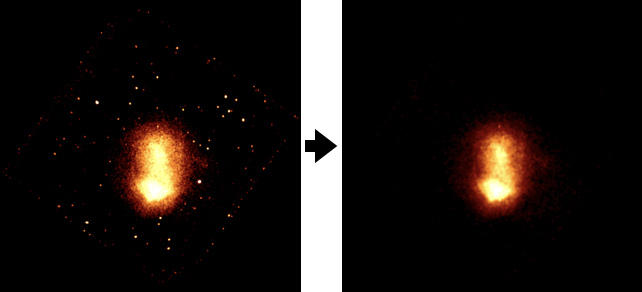}
    \caption{Both images are of the Toothbrush cluster's X-ray surface brightness. The left image includes the point sources before being excluded by the pipeline. The right image is the source-removed image used in the remaining stages of the pipeline.}
    \label{fig:stage_2}    
\end{figure*}

Stage two of the pipeline finishes with the creation of the source extracted and deflared event files. The files created (lightcurves and updated event files) can be inspected to ensure the flares are successfully removed. This stage of the process is illustrated in Figure~\ref{fig:stage_2}. The user is then prompted as to how to proceed to the next stages of the pipeline.

\subsection{Pypeline Stage 3}
\label{sec:key_stages:3}

The next stage in the pipeline is to generate the global response files for each observation. These files are integral to the spectral fitting process completed at the final stages of \texttt{ClusterPyXT}. Specifically, these are the redistribution matrix files and the auxiliary response files for each observation. The redistribution matrix files map energy space to the pulse height amplitude space as recorded by the detector. The auxiliary response files encode the quantum efficiency times the effective detector area  for each observation. This is needed to interpolate what a ``perfect" detector would see based on what was actually detected. 

\subsection{Pypeline Stage 4}
\label{sec:key_stages:4}

Stage 4 of the pipeline is relatively brief. First, the user is prompted to input a region file containing a box with all portions of the image desired to be analyzed. The final image is cropped to this region. Next, the observations and backgrounds are filtered to the 0.7-8.0 keV range. This energy range is chosen because the Chandra Proposer's Observatory Guide\footnote{\url{http://cxc.harvard.edu/proposer/POG/html/chap6.html}} indicates that the 0.7-8.0 keV range provides the optimum quantum efficiency and effective area ultimately optimizing the noise ratio. These images are then combined to be used in the next stage.

\subsection{Pypeline Stage 5}
\label{sec:key_stages:5}

The final stage of the pipeline before the spectral temperature fits is one of the most computationally intensive. This stage of the pipeline calculates the bins using an adaptive circular binning algorithm \cite{randall-2008, randall-2010}. This process generates a scale map, calculates the effective times of each region, creates the regions for each spectral fit, and prepares the scripts needed for the final stage. 

The first step is the creation of the scale map. The scale map is an image where the value at each pixel corresponds to the value of a radius.~This radius is how far out a circular region needs to be for that pixel to meet the signal to noise ratio inputted by the user. At each pixel a radius is drawn progressively larger until the  given signal to noise ratio is achieved. The periphery of the image will likely be cropped out as the edges of observations rarely contain enough data to produce a good statistical fit. 

The next step of this stage generates a region file for each observation containing circular regions for each non-zero pixel in the scale map. The radius of these regions is given by the value of the pixel in the scale map. After all regions are created, the effective observation times of each region is calculated. These times are needed to properly weight each observation. 

The last step in this stage is writing the script files which can then be used to invoke the fitting algorithm. There are multiple ways to run the final portion of these script files (see the description of the varied execution methods in the documentation).

\subsection{Spectral Fitting}\label{sec:spec_fit}
The final stage of the pipeline is the spectral fitting procedure. This operation can contain anywhere from a couple of thousand spectral fits to hundreds of thousands (maybe higher) depending on the size of the observation and resolution desired. The fitting process can be performed either in serial or parallel. In general, running the algorithm in parallel is much faster than in serial. See performance statistics in Section~\ref{sec:performance}. In order to improve the error bars in the resulting temperature map, the default fitting procedure fixes all variables other than the normalization and temperature.

The fitting procedure takes each of the ACB regions, extracts pulse invariant (PI) files from the event files, and updates the files to point to the auxiliary response files and the redistribution matrix files. \texttt{Sherpa} is used for the spectral fitting procedure \cite{sherpa}. The model used is the \texttt{X-Spec PHABS} model times the \texttt{X-Spec APEC} model. This produces a combined model with the abundance, $z$, and $N_H$ parameters fixed to the values entered by the user. The temperature (in $kT$) and the normalization parameters are then fit. While the parameters we chose to fix or fit are hard coded, a simple change to the code can allow for fixing or fitting any of the forementioned parameters. See Section~\ref{sec:model} for future plans regarding model selection. These fits and associated statistics are written to a file along with their corresponding region number. The file containing the good spectral fits is then reprojected back into an image map with the values of the fitted temperatures.

\begin{figure*}
    \centering 
    \includegraphics[width=\linewidth]{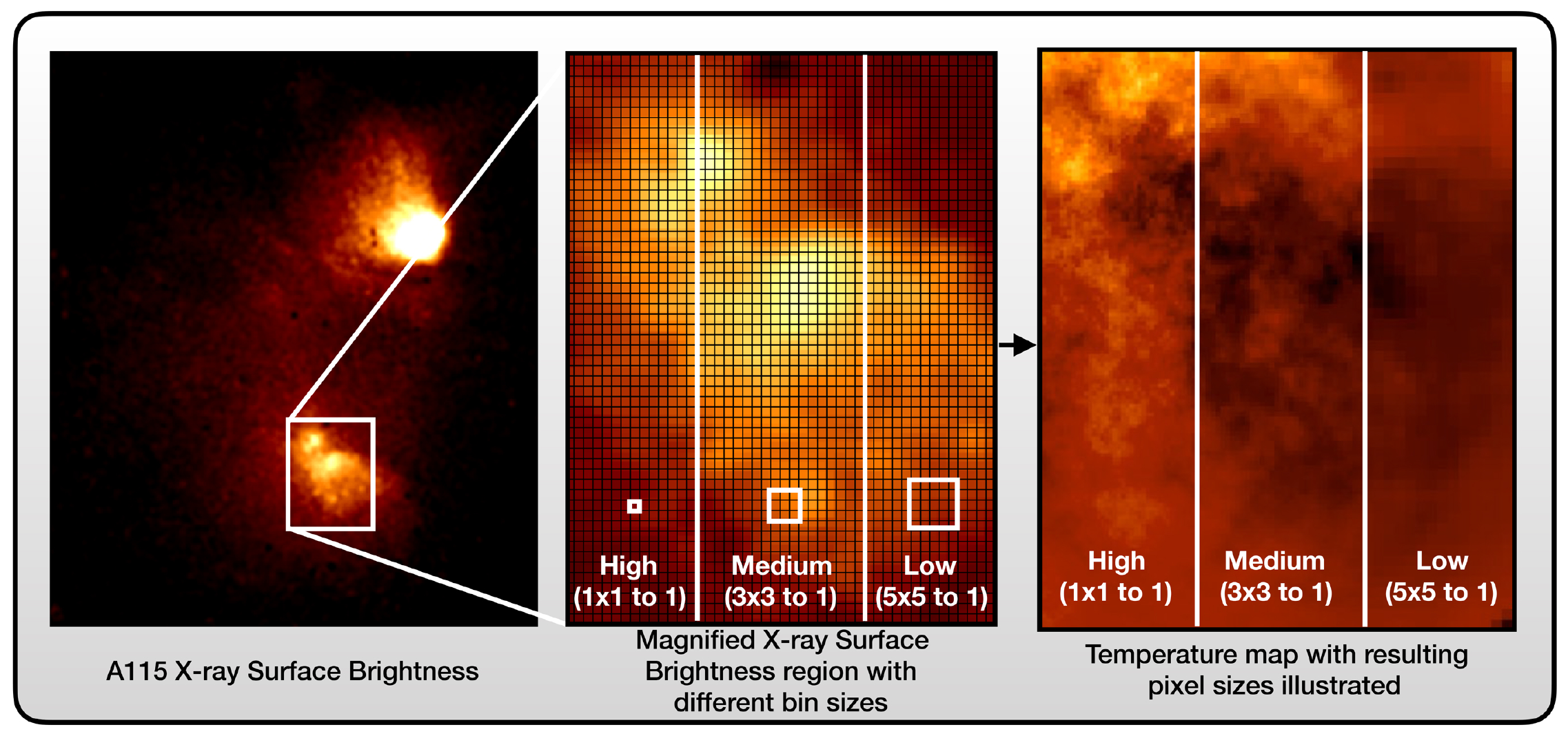}
    \caption{Depiction of the different resolution settings for the final temperature map. The left image shows an X-ray surface brightness map generated in the early stages of the pipeline. The middle panel depicts a zoomed-in region of the southern subcluster to visualize the three bin sizes available. The right image shows the resolution outcomes for each bin size: high resolution (left stripe) results in a spectral temperature fit for every pixel in the X-ray image; low resolution (right) results in essentially every fifth pixel in every fifth row being fit; if no resolution is specified, medium resolution (middle) is the default.}
    \label{fig:bin_sizes}
\end{figure*}

The spectral fitting process can be run at 3 different resolutions, low, medium, and high. The resolution setting controls how many pixels have spectral fits completed. High resolution is every pixel, medium resolution fits every other row, and low resolution every third pixel. This essentially makes the pixels $3\times3$ using the center pixel for the fit. Low resolution expands this to $5\times5$. At high resolution, minor artifacts of the binning procedure can sometimes be seen in the resulting image. A representation of the three resolutions can be seen in Figure~\ref{fig:bin_sizes}.

\section{Data Products}\label{sec:data_products}
\begin{figure*}
    \centering
    \subfloat[A2034 X-ray Surface Brightness ($S(x)$)\label{fig:A2034_xray}]{\includegraphics[width=0.48\textwidth]{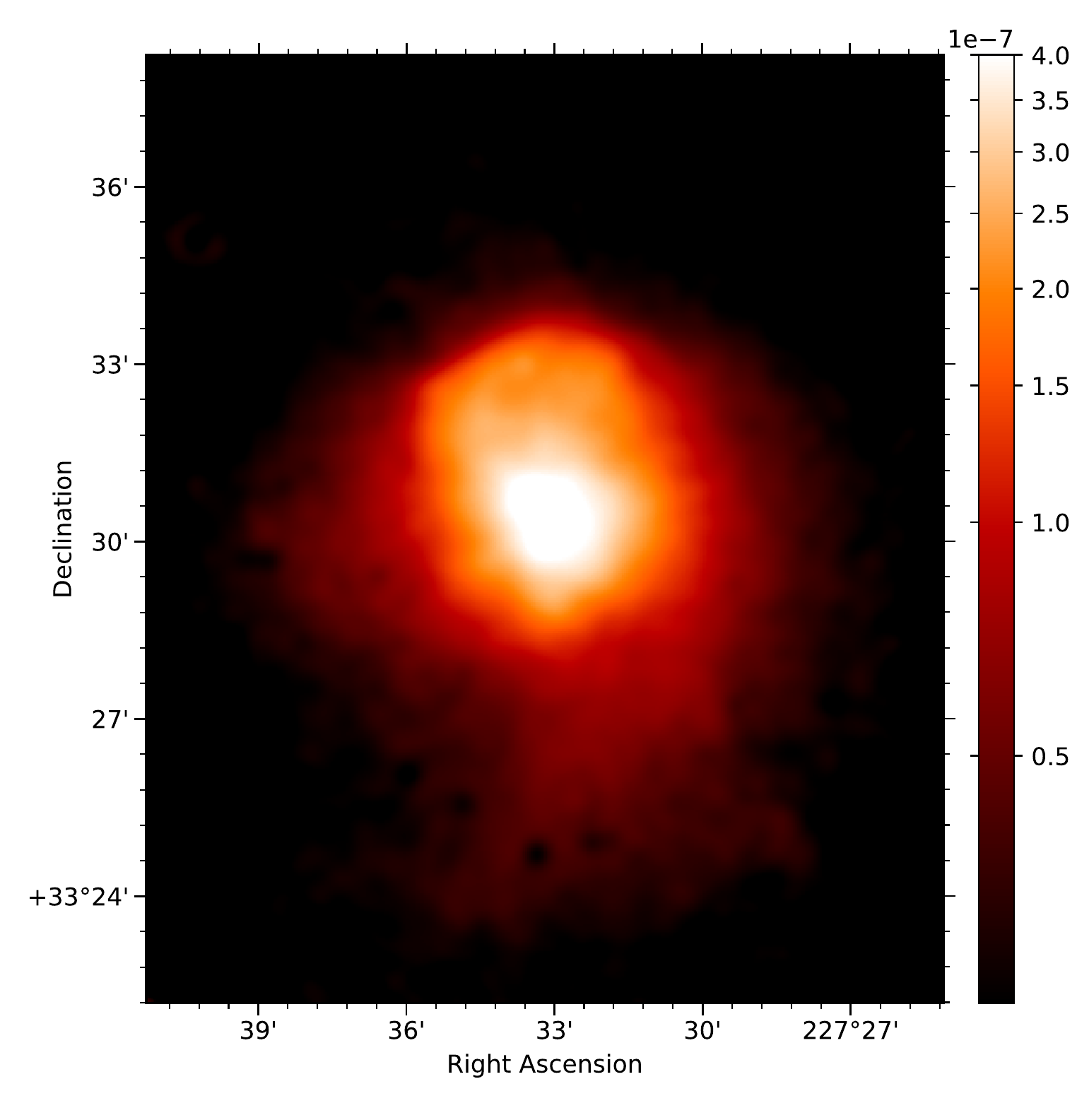}}
    \hfill
    \subfloat[A2034 Projected Pressure Map\label{fig:A2034_pressure}]{\includegraphics[width=0.48\textwidth]{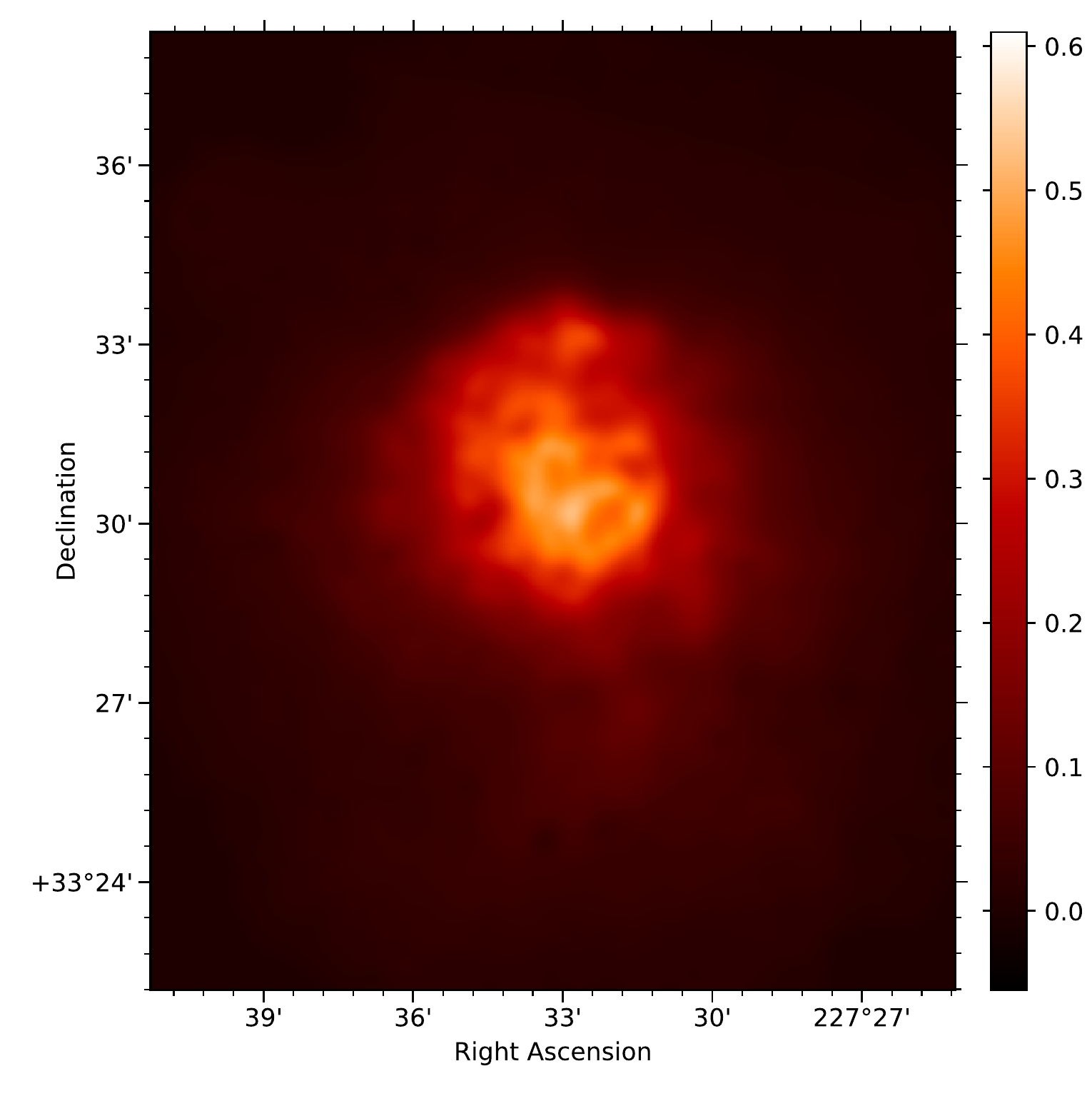}}
    \hfill
    \subfloat[A2034 Temperature Map ($T$)\label{fig:A2034_temperature}]{\includegraphics[width=0.48\textwidth]{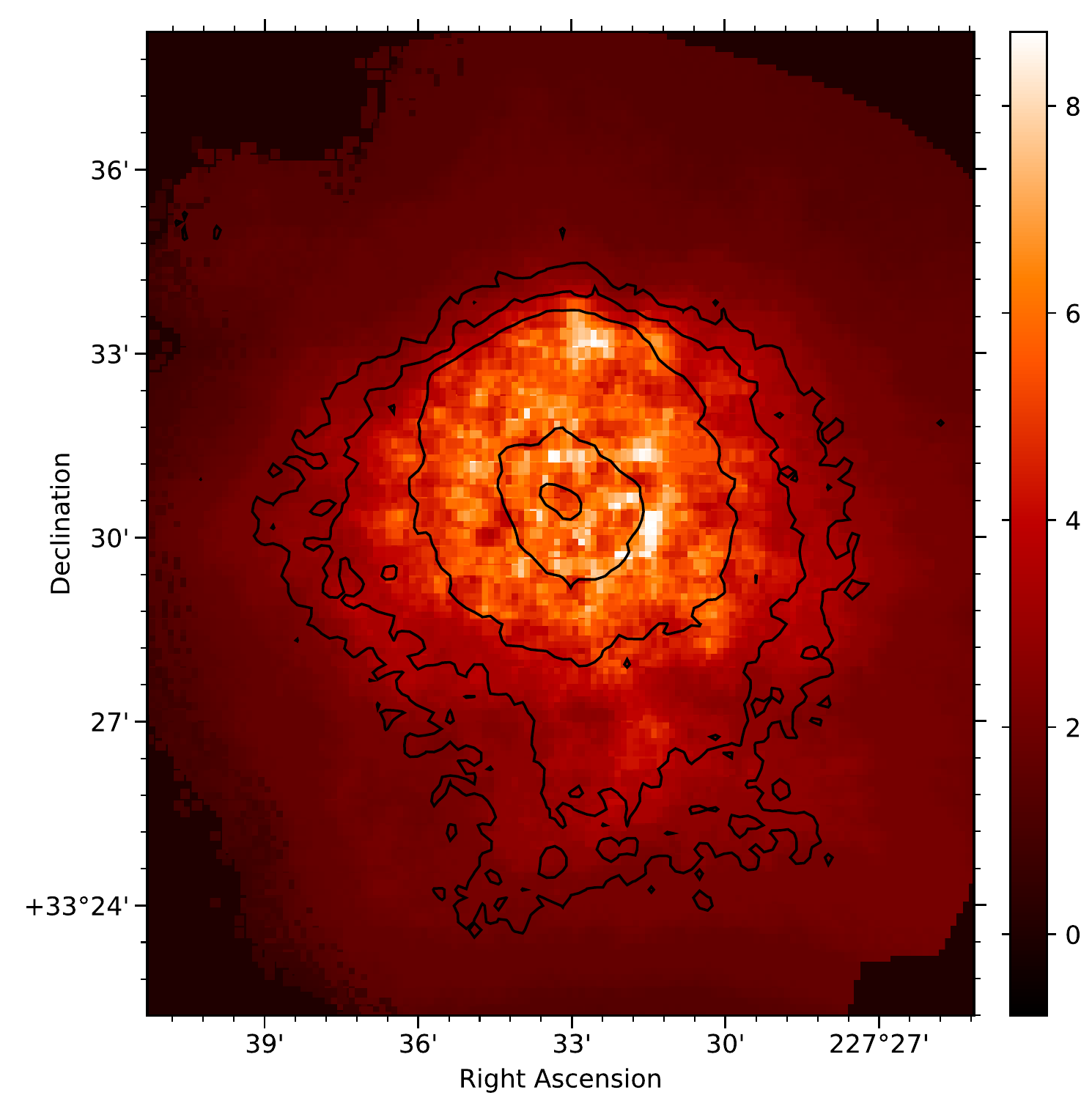}}
    \hfill
    \subfloat[A2034 Temperature Error Map\label{fig:A2034_T_Error}]{\includegraphics[width=0.48\textwidth]{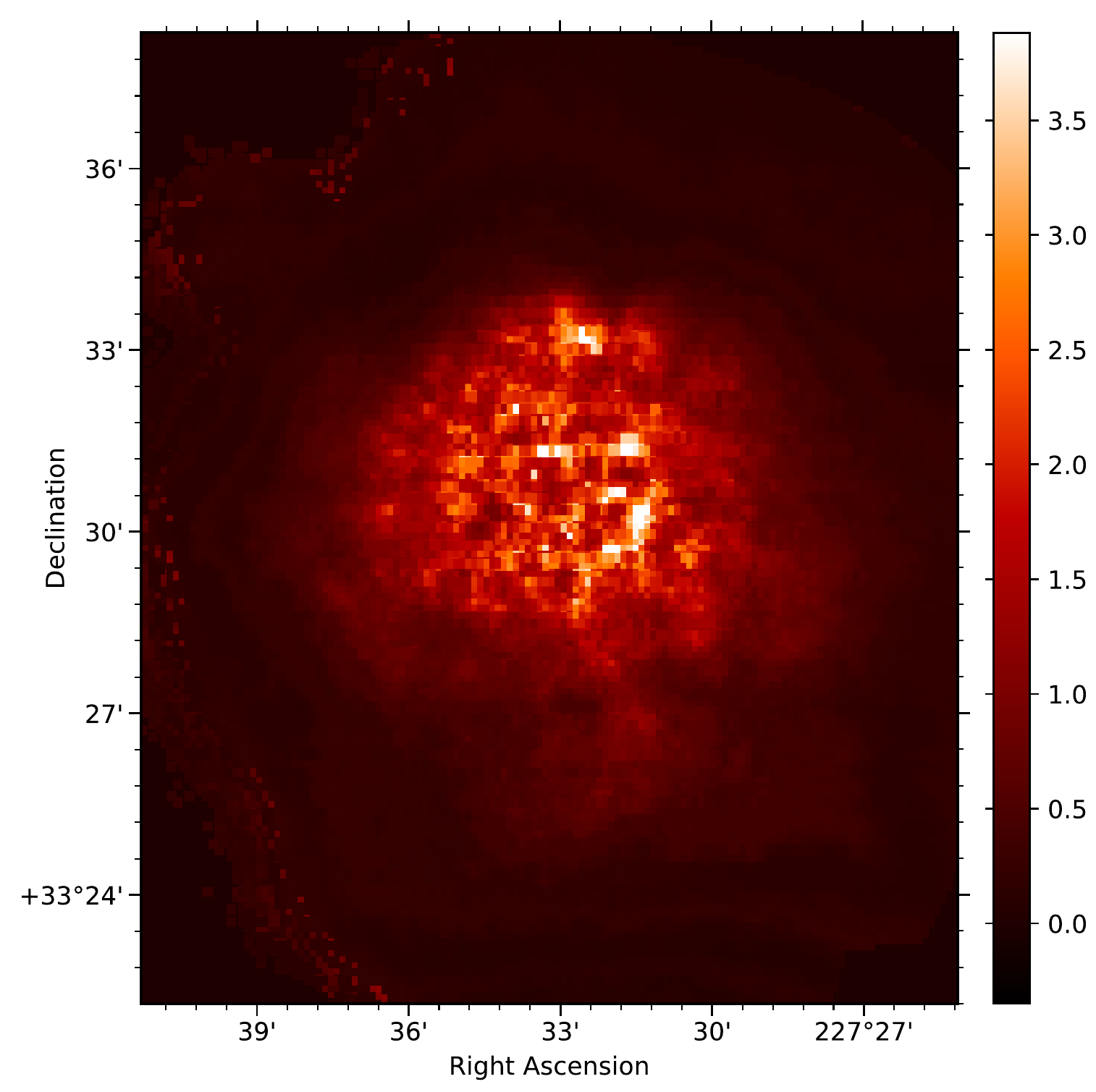}}

    \caption{ Top left: X-ray surface brightness map for A2034 (in counts/cm$^2$/s). The image is smoothed using a Gaussian kernel of 2.5$^{\prime\prime}$. The observation IDs used in this image are 7695, 2204, 12885, 13192, 12886, and 13193. Top right: Projected pseudo pressure map for A2034. This map is generated using the relation $P=[S(x)]^{(1/2)}T$. Bottom left: ACB temperature map for A2034 with X-ray surface brightness contours in black. The contour levels depicted are 4, 6, 13, 38, and 63 percent of peak emission ($7.91\times10^{-7}$ counts/cm$^2$/s). Bottom right: Temperature error map for A2034 representing 1$\sigma$ error bars for panel~\ref{fig:A2034_temperature}.}    \label{fig:A2034}
\end{figure*}

\begin{figure*}
    \centering
    \subfloat[X-ray Surface Brightness\label{fig:toothbrush_xray}]{\includegraphics[width=0.32\textwidth]{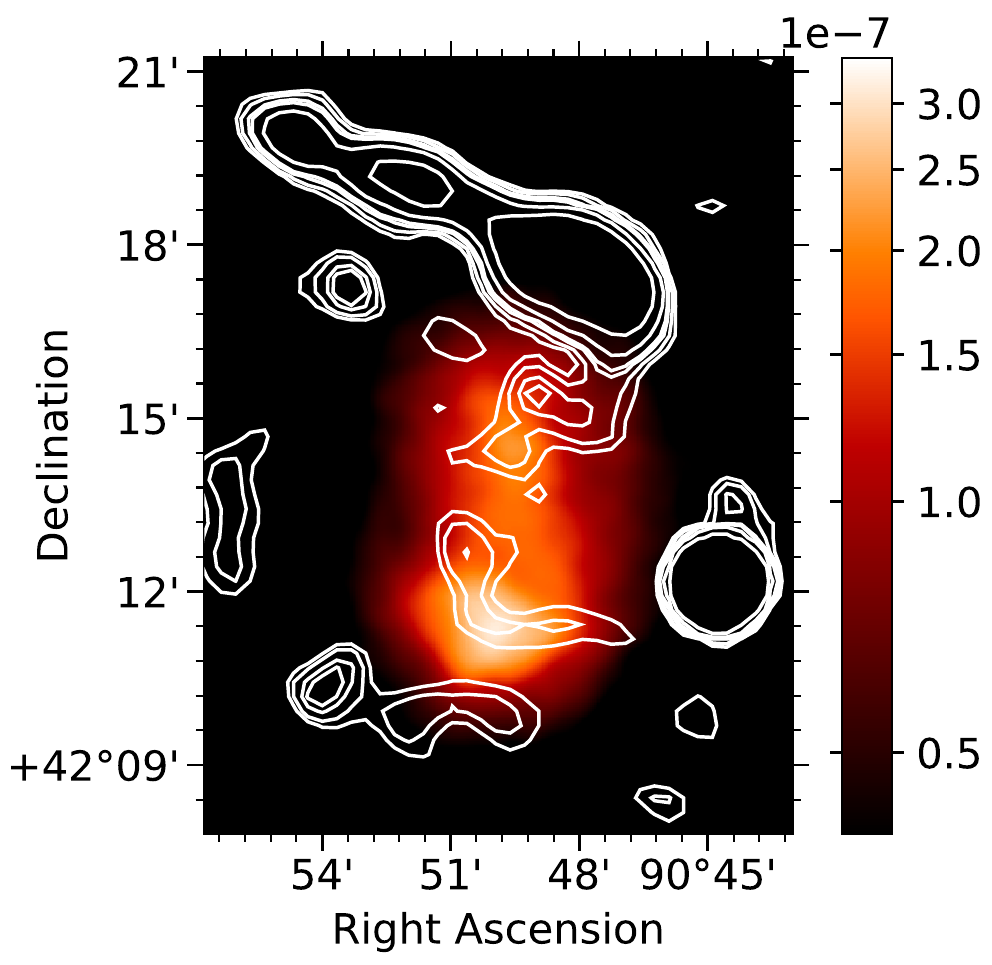}}
    \hfill
    \subfloat[Temperature Map\label{fig:toothbrush_temperature}]{\includegraphics[width=0.3\textwidth]{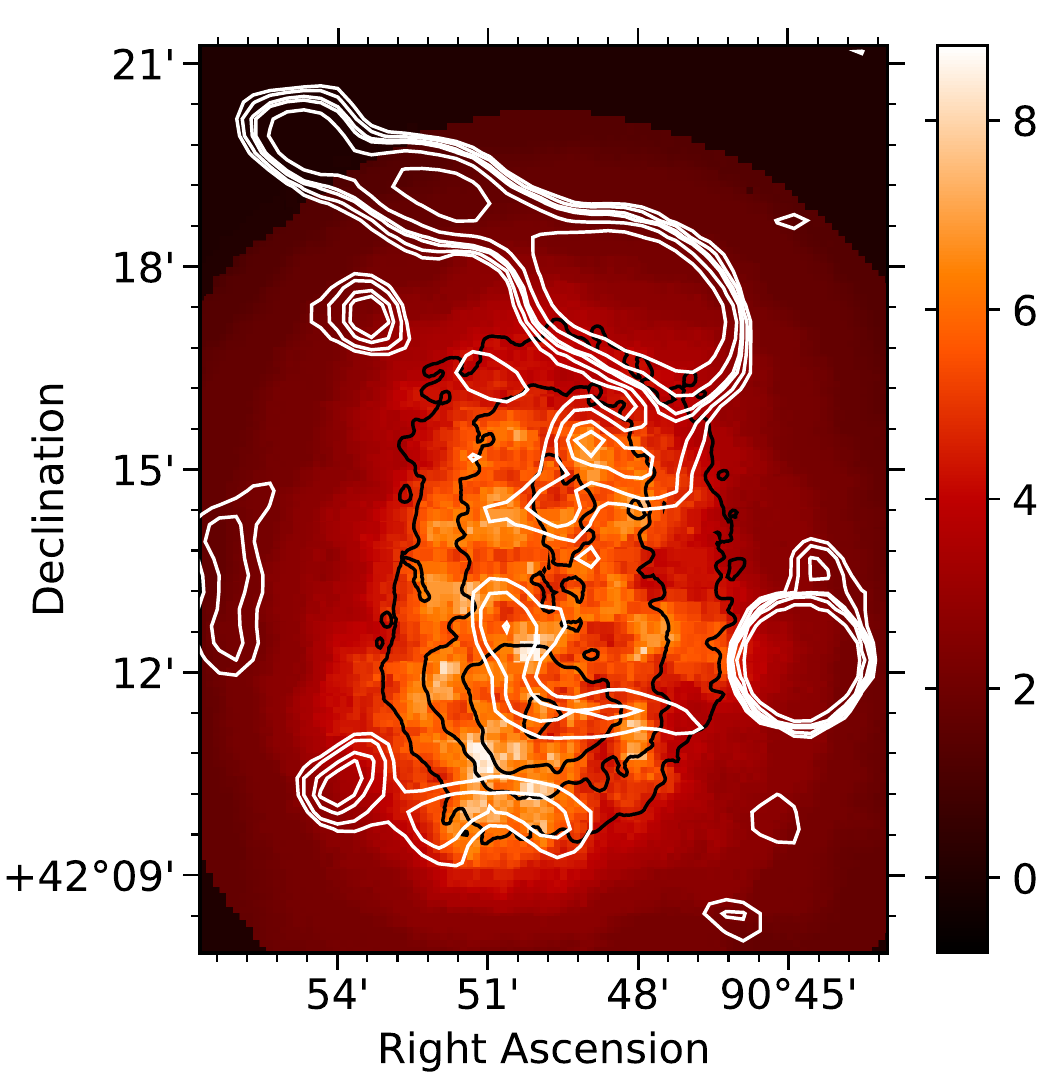}}
    \hfill
    \subfloat[Pressure Map\label{fig:toothbrush_pressure}]{\includegraphics[width=0.315\textwidth]{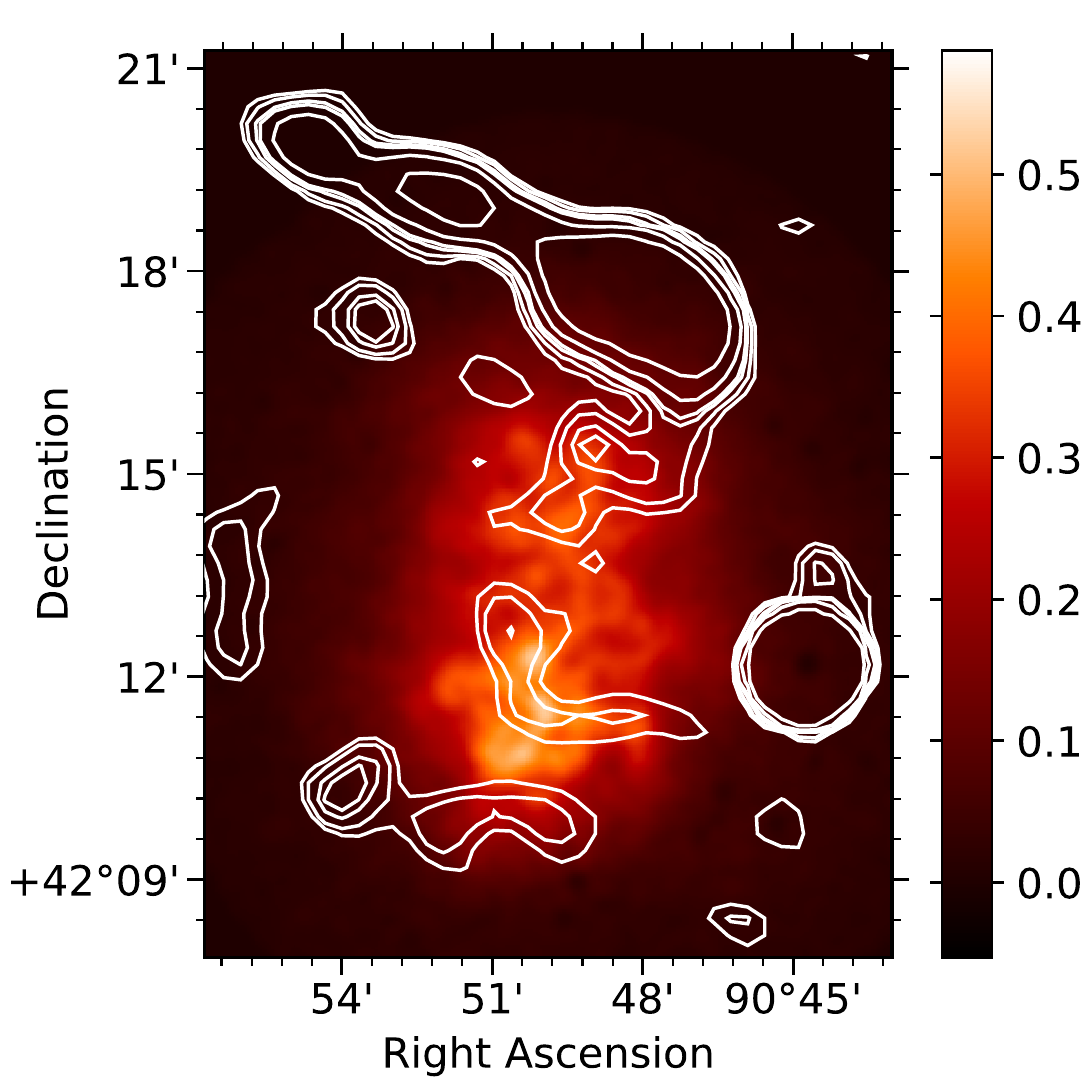}}
    \caption{ All: White contours represent 1.4 GHz radio emission. Radio data from the National Radio Astronomy Observatory's (NRAO) Very Large Array (VLA) Sky Survey (NVSS) \cite{nvss}. The radio relic, from which the cluster receives its nickname `Toothbrush', is the linear feature in the northern part of the images. It is depicted with contour levels at 1.5, 2.25, 3.3, 5, 6.6, and 13.2 percent of peak emission ($\sim 7.6\times10^{-2}$ Jy/Beam) \cite{nvss}. Left: X-ray surface brightness map of the Toothbrush cluster in units of counts/cm$^2$/s. The image is smoothed with a 2.5$^{\prime\prime}$ Gaussian kernel. The observation IDs used in this image are 15171, 15172, and 15323. Center: Temperature map of the Toothbrush cluster with X-ray surface brightness contours in black. X-ray contour levels are 14, 27, 54, and 87 percent of the peak emission ($3.47\times10^{-7}$ counts/cm$^2$/s). Right: Relative projected pseudo pressure map of the Toothbrush cluster.
    }
    \label{fig:toothbrush}
\end{figure*}

While the main output of the pipeline is the ACB temperature map, multiple files are generated along the way which may be useful for further analysis. This includes a projected pressure map, a combined X-ray surface brightness map, and in the future, an entropy map. The creation of the X-ray surface brightness and temperature maps is outlined above in Section~\ref{sec:spec_fit}. To create the pressure map, the X-ray surface brightness and temperature maps are combined. As surface brightness is proportional to $n^2$ and $P=nk_BT$, the square root of the surface brightness map can be multiplied by the temperature map to obtain a relative pressure map. This combination is achieved by using the \texttt{dmimgcalc} command with the surface brightness and temperature maps as operands. Examples of these data products can be seen in Figures \ref{fig:A2034} and \ref{fig:toothbrush}.

\section{Run-time Performance}
\label{sec:performance}

Run-time performance is highly dependent on the hardware running the code. While the spectral fitting portion of the pipeline is best suited for high-end server class hardware or even supercomputers (such as NASA Ames Pleiades), the bulk of the code is suited for consumer grade hardware. An Apple Macbook Air from 2013, with a Intel Core i7 1.7GHz processor and 8GB of ram processes a cluster up to the fitting portion of the pipeline in under 24 hours with little user interaction. This is highly dependant on the number and exposure time of the observations. A small cluster with less than a hundred kiloseconds of total exposure time can generally be processed in under 12 hours. The Coma cluster, with over 500 kiloseconds of exposure time will likely run into the upper limit of 24 hours. 

The spectral fitting portion of the pipeline is best suited for a server-class desktop or a supercomputer. The number of spectral fits required to produce a temperature map is on order of 10$^4$ to 10$^5$ depending upon the resolution selected. These fits generally take around 0.1-2 minutes per fit depending upon the amount of observational data. This gives a wide range of possible compute times (0.5-140 computing days). Spectral fitting a small cluster with less than a hundred kiloseconds of observations can be achieved on a consumer grade iMac (common in universities) in under a day running in parallel across 4 cores.  

\section{Future development}

There are many areas for which further development is planned or would be desired. The list below is by no means inclusive and requests to the GitHub page are welcome. Most of these enhancements are to extend the pipeline beyond the production of temperature maps only for Chandra observations. While development is continuous, this is an open source project and contribution is strongly encouraged. Please check the GitHub repository for an updated list. 

\subsection{User Interface}

The current user interface is command line driven and offers basic functionality. The enhancement proposed here is providing a full menu system. As of right now, the user needs to edit the configuration files for each cluster in order to rerun any portion of the pipeline. Ideally this interface would allow the user to jump backwards to any stage of the pipeline previously completed, along with allowing editing of the configuration files and fitting parameters from within the interface itself, without an external editor required.

\subsection{XMM-Newton}
\label{sec:xmm}

Adding the ability to process XMM-Newton X-ray observations to the pipeline can help increase the fidelity and quality of the pipeline if integrated properly. Ideally XMM observations would be input alongside Chandra observations in the cluster initialization stage.~The XMM observations could then be used together with those from Chandra to create data products using the combined observations.

\subsection{Expand Multiprocessor Support}

While the most computationally expensive portion of the pipeline ---spectral fitting--- supports multiprocessing, there are other areas ripe for such an implementation. Both the scale map generation and effective time calculations could see orders of magnitude speed improvements with multiple cores processing the data at the same time. Further exploration needs to be carried out into supporting GPU execution.

\subsection{User Interface Based Model Selection}\label{sec:model}
The initial release of \texttt{ClusterPyXT} has the metallicity, redshift, and hydrogen column density parameters fixed with the fitting parameters being temperature and normalization. Future iterations will include the ability to change which parameters are fixed and which are fit to the data. This will allow the user to generate additional data products such as metallicity maps in addition to temperature maps.

\section{Conclusion}
X-ray spectral temperature maps can provide insight into the physical processes occurring within a galaxy cluster. The data products produced by \texttt{ClusterPyXT} can help highlight areas of turbulence, shocks, and dynamic activity, which in turn can lead to a better understanding of the underlying physics behind these phenomena as well as those of radio relics and halos.

\texttt{ClusterPyXT} automates the creation of X-ray temperature, pressure, surface brightness and density maps. This software downloads, merges, filters, and processes X-ray observations to a point ready for spectral fitting with little user interaction. The temperature maps produced allow for a high-resolution, qualitative view of the galaxy cluster state. Analyses of the physics occurring within a galaxy cluster can then be enhanced with the data products produced by \texttt{ClusterPyXT}.

\section{Acknowledgements}
The original code for a pipeline that builds temperature maps was created in IDL by Greg Salvesen using an alternative binning method. The ACB method and shock finding routines were written by Abhi Datta and David Schenck in Bash script. Later on, these scripts were made more efficient and parallelized by Chris Yarrish.

This research was supported by NASA ADAP grant NNX15AE17G. BA is supported by the US Department of Veterans Affairs Vocational Rehabilitation and Employment program. DR is supported by a NASA Postdoctoral Program Senior Fellowship at NASA's Ames Research Center, administered by the Universities Space Research Association under contract with NASA.

%\clearpage
%\section{References}
%\bibliographystyle{plain}
\bibliography{bib}
\end{document}